\documentclass[%
reprint,
superscriptaddress,
 amsmath,amssymb,
 aps,
]{revtex4-2}

\usepackage[utf8]{inputenc}
\usepackage{graphicx}
\usepackage{dcolumn}
\usepackage{multirow}
\usepackage{bm}
\usepackage{mathrsfs}
\usepackage{hyperref}
\hypersetup{
  colorlinks = true,
  urlcolor = blue,
  linkcolor = blue,
  citecolor = green,
  filecolor = magenta,
}

\newcommand{\ud}{\mathrm{d}}
\newcommand{\pd}{\partial}

\newcommand{\lie}{\mathscr L}

\newcommand{\order}[1]{\mathcal O\left(#1\right)}

\newcommand{\ma}{\mathfrak a}
\newcommand{\mb}{\mathfrak b}

\begin{document}


\title{Asymptotic analysis of Chern-Simons modified gravity and its memory effects}

\author{Shaoqi Hou}
\email{hou.shaoqi@whu.edu.cn}
\affiliation{School of Physics and Technology, Wuhan University, Wuhan, Hubei 430072, China}
\author{Tao Zhu}
\email{zhut05@zjut.edu.cn}
\affiliation{Institute for theoretical physics and cosmology, Zhejiang University of Technology,Hangzhou, Zhejiang 310032, China}
\author{Zong-Hong Zhu}
\email{zhuzh@whu.edu.cn}
\affiliation{School of Physics and Technology, Wuhan University, Wuhan, Hubei 430072, China}
\affiliation{Department of Astronomy, Beijing Normal University, Beijing 100875,  China}

\date{\today}

\begin{abstract}
    We study the asymptotically flat spacetime in Chern-Simons modified gravity, and then 
    gravitational memory effects  are considered in this work.
    If the Chern-Simons scalar does not directly couple with the ordinary matter fields, there are also displacement, spin and center-of-mass memory effects as in general relativity.
    This is because the term of the action that violates the parity invariance is linear in the scalar field but quadratic in the curvature tensor.
    This results in the parity violation occurring at the higher orders in the inverse luminosity radius.
    Although there exists the Chern-Simons scalar field, interferometers, pulsar timing arrays, and the Gaia mission are incapable of detecting its polarization, 
    so the scalar field does not induce any new memory effects that can be detected by interferometers or pulsar timing arrays.
    The asymptotic symmetry group is also the extended Bondi-Metzner-Sachs group.
    The constraints on the memory effects excited by the tensor modes are obtained as well.
\end{abstract}

\maketitle


\section{Introduction}

The gravitational memory effect is an intriguing phenomenon, which often refers to the lasting change in the relative distance between test particles after the gravitational wave disappears.
This effect, sometimes named displacement memory, was first discovered in general relativity nearly 50 years ago \cite{Zeldovich:1974gvh,Braginsky:1986ia,Christodoulou1991,Thorne:1992sdb}.
Recently, new memory effects were identified, such as spin memory \cite{Pasterski:2015tva,Mao:2018xcw}, and center-of-mass (CM) memory \cite{Nichols:2018qac}.
They contribute to the so-called subleading displacement memory, the permanent change in the relative distance when the initial relative velocity is nonzero.
The spin memory effect also leads to the difference in the periods of two counterorbiting massless particles in a circular orbit \cite{Pasterski:2015tva}.
Finally, there also exists the velocity memory effect, which is just the lasting change in the relative velocity between test particles \cite{Zhang:2017rno,Zhang:2017geq,Zhang:2018srn,Compere:2018ylh,Mao:2019sph}.
All of the above effects are characterized by the permanent change in certain physical quantities, and similar phenomena also occur in electromagnetism and Yang-Mills theory and more \cite{Dinu:2012tj,Bieri:2013hqa,Strominger:2018inf}.
In this work, we will focus on the memories in gravitation.

Memory effects are deeply related to the asymptotic symmetries of the spacetime. 
In the case of the asymptotically flat spacetime, the memory effects usually considered take place near the (future) null infinity \cite{Wald:1984rg}, at which null geodesics eventually arrive.
The asymptotic symmetries are diffeomorphisms that, roughly speaking, preserve the geometry of the null infinity \cite{Geroch1977}.
Studies have shown that the asymptotic symmetries include the supertranslation symmetries and the Lorentz symmetries \cite{Bondi:1962px,Sachs:1962zza}.
They together form the so-called Bondi-Metzner-Sachs (BMS) group, which is an infinite dimensional group, generalizing the Poincar\'e group. 
Supertranslations are certain generalizations of the usual translations, while Lorentz transformations are actually the conformal transformations on a unit 2-sphere generated by the global conformal Killing vector fields.
New works have extended the BMS group by allowing the conformal Killing vectors to have isolated singularities on the 2-sphere \cite{Barnich:2009se,Barnich:2010eb} or by replacing Lorentz transformations by all of the diffeomorphisms on the 2-sphere \cite{Campiglia:2014yka,Campiglia:2015yka}.
The former gives rises to the extended BMS group, and the latter might suffer from the diverging symplectic current \cite{Flanagan:2019vbl}, so it will not be considered in this work.

Because of the supertranslation symmetry, there are infinitely many degenerate vacuum states in the gravity sector that can be transformed into each other via the supertranslation transformations.
The vacuum transition explains the displacement memory effect \cite{Strominger:2014pwa}.
The displacement memory is also constrained by the flux-balance laws associated with the supertranslations.
Similar constraints can be found for spin and CM memories. 
In particular, flux-balance laws associated with the superrotations \footnote{Here, superrotations specifically refer to the generalizations of spatial rotations in the extended BMS group.} constrain the spin memory effect, while flux-balance laws with the superboosts \footnote{Here, superboost specifically refer to the generaliations of Lorentz boosts in the extended BMS group.} constrain the CM memory effect.
These flux-balance laws play important roles in the determination of the strength of the memory effect and thus its observation.
The detection of gravitational waves \cite{Abbott:2016blz,TheLIGOScientific:2017qsa,LIGOScientific:2018mvr,LIGOScientific:2020ibl} not only proved the existence of gravitational waves, but also made it possible to observe the memory effect.
Indeed, ground-based interferometers such as aLIGO could detect memories \cite{Lasky:2016knh,McNeill:2017uvq,Johnson:2018xly,Hubner2020mmn,Boersma:2020gxx,Zhao:2021hmx}. 
They can also be measured by pulsar timing arrays \cite{Seto:2009nv,Wang2015mm} and the Gaia mission \cite{Madison:2020xhh}.
The spin memory effect is observable by LISA \cite{Pasterski:2015tva}, but the CM memory is more difficult to be detected with the current and even the planed detectors \cite{Nichols:2018qac}.

This is roughly what has happened in general relativity.
Although general relativity is a very successful theory \cite{Will:2014kxa}, it still suffers from some problems, such as its breakdown at the singularity, the nonrenormalizability, dark matter and dark energy, etc. 
To resolve at least some of these problems, there have been a plethora of modified theories proposed.
Among them, Brans-Dicke theory \cite{Brans:1961sx} is the simplest, which contains one extra gravitational degree of freedom, the Brans-Dicke scalar field.
Memory effects of this theory have been studied in Refs.~\cite{Hou:2020tnd,Tahura:2020vsa,Hou:2020wbo,Seraj:2021qja,Tahura:2021hbk}.
It was found out that in addition to the memories already discovered in general relativity, there also exits the one excited by the Brans-Dicke scalar, dubbed S memory \cite{Du:2016hww}.
These memories are also related to the asymptotic symmetries, and constrained by the corresponding flux-balance laws, although S memory is more subtle \cite{Seraj:2021qja}.
Very interestingly, the S memory effect can also be used to distinguish general relativity from Brans-Dicke theory \cite{Du:2016hww,Koyama:2020vfc}.
One thus speculates that the study of the memory effect in modified theories of gravity may help probe the nature of gravity.

In this work, memory effects in a different modified gravity theory, Chern-Simons modified gravity \cite{Jackiw:2003pm,Alexander:2009tp}, will be studied.
This theory also includes one extra gravitational  degree of freedom, called the Chern-Simons scalar field, but it is a pseudoscalar.
Thus the effects of parity violation might take place.
For example, in the cosmological background, the gravitational wave might experience the amplitude and the velocity birefringences as predicted in a generic parity violating theory which incorporates Chern-Simons gravity as a special case  \cite{Qiao:2019wsh,Zhao:2019xmm,Kamada:2021kxi}.
Indeed, Chern-Simons gravity predicts that the left-handed and the right-handed (tensor) gravitational waves propagate at different damping rates -- the amplitude birefringence -- but they both travel at the speed of light.

With the Bondi-Sachs formalism \cite{Bondi:1962px,Sachs:1962wk,Madler:2016xju}, one can obtain the metric and the scalar fields in the asymptotically flat spacetime in this theory. 
We show that the metric resembles the one in general relativity with a canonical scalar field at the lower orders in the inverse of the luminosity radius.
The parity violating terms explicitly appear at the higher orders.
The asymptotic symmetries are thus expected to be the same as the extended BMS symmetries in general relativity \cite{Flanagan:2015pxa}.
There are also the same memories induced by the tensor degrees of freedom.
Since the Chern-Simons scalar field is assumed not to couple with the matter fields, interferometers, pulsar timing arrays, and the Gaia mission are not capable of detecting its memory effects even if they exist.
Although in this work, the conserved charges and fluxes will not be determined, the constraints on memory effects can still be obtained with the equations of motion.

This work is organized in the following way. 
Section~\ref{sec-cs} briefly reviews Chern-Simons modified gravity.
Section~\ref{sec-afs} focuses on the asymptotically flat spacetime of this theory.
In particular, one first discusses the boundary conditions that the metric and the scalar fields should satisfy in Sec.~\ref{sec-bc}, so that the asymptotic solutions can be determined in Sec.~\ref{sec-asol}.
After that, the asymptotic symmetries are obtained in Sec.~\ref{sec-bms}.
Then, memory effects are discussed in Sec.~\ref{sec-mm}.
These effects can be introduced via solving the geodesic deviation equations in Sec.~\ref{sec-gde}.
Then memories are related to the vacuum transition in Sec.~\ref{sec-vt}.
Section~\ref{sec-cmm} presents the constraints on memory effects by integrating the equations of motion.
Finally, a brief summary \ref{sec-con} concludes this work.
Throughout this paper, $c=1$.
Most of the calculation was done with the help of \verb+xAct+ \cite{xact}.

\section{Chern-Simons modified gravity}
\label{sec-cs}

The action of Chern-Simons modified gravity is \cite{Alexander:2009tp}
\begin{equation}
    \label{eq-cs-act}
    \begin{split}
    S=\int\ud^4x\sqrt{-g}\bigg(&\kappa R+\frac{\mathfrak a}{4}\vartheta R_{abcd}{}^*R^{bacd}\\
    &-\frac{\mathfrak b}{2}\nabla_a\vartheta\nabla^a\vartheta-\mathfrak b V(\vartheta)\bigg)+S_m,
    \end{split}
\end{equation}
where $\kappa=1/16\pi G$, and $\mathfrak a$ and $\mathfrak b$ are all coupling constants.
$S_m$ is the action for matter fields and does not depend on $\vartheta$.
$V(\vartheta)$ is the potential for the Chern-Simons scalar $\vartheta$. 
Here, we consider the special case with $V(\vartheta)=0$ so that $\vartheta$ is massless.
Then, the action acquires the shift symmetry under the addition of a constant to $\vartheta$.
${}^*R^{bacd}=\epsilon^{cdef}R^{ba}{}_{ef}/2$ is the Hodge dual.
The second term in the action probably arises  due to the gravitational anomaly of the standard model of elementary particles \cite{Fujikawa:1979ay,Fujikawa:1980eg}, the Green-Schwarz anomaly canceling mechanism in string theory \cite{Polchinski:1998rr}, or the scalarization of the Barbero-Immirzi parameter in loop quantum gravity \cite{Perez:2005pm,Randono:2005up}.
Because of the presence of $\epsilon^{abcd}$, $\vartheta$ is a pseudoscalar in order that the action $S$ is invariant under the parity transformation.
If one ignores the second term in the action, one obtains general relativity with a canonical scalar field $\sqrt{\mb}\vartheta$.
In this work, we will not consider the matter action for simplicity.
This also causes us to assume the matter fields decay sufficiently fast as the distance to the source is approaching infinity.

The equations of motion are \cite{Alexander:2009tp}
\begin{gather}
    R_{ab}-\frac{1}{2}g_{ab}R+\frac{\mathfrak a}{\kappa}C_{ab}=\frac{1}{2\kappa}T_{ab}^{(\vartheta)},\label{eq-ein}\\
    \nabla_a\nabla^a\vartheta=-\frac{\mathfrak a}{4\mathfrak b}R_{abcd}{}^*R^{bacd}.\label{eq-eom-th}
\end{gather}
Here, $C_{ab}$ is called the C-tensor, given by 
\begin{equation}
    \label{eq-def-ct}
    C^{ab}=(\nabla_c\vartheta)\epsilon^{cde(a}\nabla_eR^{b)}{}_d+(\nabla_c\nabla_d\vartheta){}^*R^{c(ab)d},
\end{equation}
where $\nabla_a\vartheta$ and $\nabla_a\nabla_b\vartheta$ are also called the Chern-Simons velocity and acceleration, respectively \cite{Alexander:2008wi}.
$T_{ab}^{(\vartheta)}$ is the stress energy tensor of the Chern-Simons scalar,
\begin{equation}
    \label{eq-def-ts}
    T^{(\vartheta)}_{ab}=\mathfrak b\left(\nabla_a\vartheta\nabla_b\vartheta-\frac{1}{2}g_{ab}\nabla_c\vartheta\nabla^c\vartheta\right).
\end{equation}
Since $\ma$ and $\mb$ are free, one may set $\ma\ne0$ and $\mb=0$. 
Then $\vartheta$ should be prescribed by hand, and one is now in the  \emph{nondynamical}  framework.
When neither $\ma$ nor $\mb$ is zero, $\vartheta$ has its own dynamics. 
This framework is called \emph{dynamical}. 
In this paper, we work in the dynamical framework.

Chern-Simons gravity has many applications in astrophysics, cosmology, and so on.
Therefore, it is constrained by astrophysical tests, solar system tests, and cosmological observations.
For example, the energy scale  above which the parity is violated arising from the Chern-Simons gravity is at least $10^{-3}\text{ km}^{-1}$ from the observation of the torque-induced precession in the solar system \cite{Smith:2007jm}.
The energy scale beyond which the parity violation effect is strong was found to be at least 33 meV from the binary pulsar observation \cite{Yunes:2008ua}.
For more phenomenology and constraints, please refer to Ref.~\cite{Alexander:2009tp}.

\section{Asymptotically flat spacetimes}
\label{sec-afs}

Roughly speaking, the asymptotically flat spacetime is the one approaching the Minkowski spacetime at distances very far away from the source of the gravity.
In the relativistic theory, there are three types of ways to approach the infinity: along timelike, spacelike, or null directions.
For problems involving (massless) radiation, it is useful to consider the spacetime region near the null infinity approached by null geodesics.
In general relativity, one can define the so-called asymptotically flat spacetime at the null infinity using the conformal completion technique \cite{Penrose:1962ij,Penrose:1965am,Wald:1984rg}.
Or, one may also be able to impose certain asymptotic behaviors of the metric or other fields near the null infinity in a suitable coordinate system, usually, Bondi-Sachs coordinates $(u,r,x^2=\theta,x^3=\phi)$ \cite{Bondi:1962px,Sachs:1962wk,Barnich:2010eb},
\begin{equation}
    \label{eq-bc}
    \begin{split}
    \ud s^2=&e^{2\beta}\frac{V}{r}\ud u^2-2e^{2\beta}\ud u\ud r\\
    &+h_{AB}(\ud x^A-U^A\ud u)(\ud x^B-U^B\ud u),
    \end{split}
\end{equation}
where $\beta,\,V,\,U^A$, and $h_{AB}$ ($A,B=2,3$) are six metric functions.
One can similarly define the asymptotically flat spacetime in modified theories of gravity, for example, Brans-Dicke theory \cite{Brans:1961sx} as done in Refs.~\cite{Hou:2020tnd,Tahura:2020vsa,Hou:2020wbo}.
Here, for simplicity, we will assign suitable asymptotic behaviors to the metric $g_{ab}$ and the Chern-Simons scalar $\vartheta$ in Bondi-Sachs coordinates in order to define the asymptotic flatness in Chern-Simons gravity.

\subsection{Boundary conditions}
\label{sec-bc}

In general relativity, the boundary conditions for the metric field in the asymptotically flat spacetime at null infinity are given by \cite{Bondi:1962px,Sachs:1962wk,Barnich:2010eb}
    \begin{gather}
    g_{uu}=-1+\order{r^{-1}},\label{eq-guu-bdy}\\ g_{ur}=-1+\order{r^{-2}},\\ g_{uA}=\order{1},\\
    g_{rr}=g_{rA}=0,\\ h_{AB}=r^2\gamma_{AB}+\order{r},\label{eq-h-bdy}
    \end{gather}
where  $\gamma_{AB}$ is the round metric on a unit 2-sphere, 
\begin{equation}
    \label{eq-gmet}
    \gamma_{AB}\ud x^A\ud x^B=\ud\theta^2+\sin^2\theta\ud\phi^2.
\end{equation}
In addition, the determinant of $h_{AB}$ is required to be 
\begin{equation}
    \label{eq-det-gr}
    \det(h_{AB})=r^4\sin^2\theta,
\end{equation}
so $r$ is the luminosity radius \cite{Bondi:1962px,Sachs:1962wk}.
In terms of the metric functions, one can find out that \cite{Barnich:2010eb}
\begin{equation*}
    \label{eq-exp-m}
    \beta=\order{r^{-1}},\quad V=-r+\order{r^0},\quad U^A=\order{r^{-2}}.
\end{equation*}
For Chern-Simons gravity, one may propose different boundary conditions.
However, by examining Einstein's equations \eqref{eq-ein}, one realizes that if one ignores the C-tensor term, one knows that the equations describe a spacetime sourced by a canonical scalar field, just as in general relativity.
Since near the null infinity, the spacetime resembles the flat one, any deviation from the Minkowski metric can be treated as the small perturbation.
Then for the purpose of determining the boundary conditions, it might be reasonable to ignore the C-tensor term as it represents a higher order term.
Therefore, we just impose the same boundary conditions on $g_{ab}$ as in general relativity.
In addition, one also requires that $\vartheta=\vartheta_0+\order{1}$.
The chosen boundary conditions also imply that the parity violating effects would be of the higher orders as shown below.

Now, the asymptotic behaviors can be written down.
One expands the metric functions in the following way \cite{Flanagan:2015pxa},
\begin{gather}
    \beta=\frac{\beta_1}{r}+\frac{\beta_2}{r^2}+\frac{\beta_3}{r^3}+\order{\frac{1}{r^4}},\label{eq-exp-beta}\\
    \begin{split}
    U^A=&\frac{\mathcal U^A}{r^2}+\frac{1}{r^3}\left[ -\frac{2}{3}N^A+\frac{1}{16}\mathscr D^A(c_{BC}c^{BC})\right.\\
    &\left.+\frac{1}{2}c^{AB}\mathscr D^Cc_{BC} \right]+\frac{\mathscr U^A}{r^4}+\order{\frac{1}{r^5}},\label{eq-user}
    \end{split}\\
    V=-r+2m+\frac{2\mathcal M}{r}+\order{\frac{1}{r^2}},\\
    h_{AB}=r^2\gamma_{AB}+rc_{AB}+d_{AB}+\frac{e_{AB}}{r}+\order{\frac{1}{r^2}}.\label{eq-exp-h}
\end{gather}
Similarly, one has 
\begin{equation}
    \label{eq-thex}
    \vartheta=\vartheta_0+\frac{\vartheta_1}{r}+\frac{\vartheta_2}{r^2}+\order{\frac{1}{r^3}}.
\end{equation}
In the above expansions, all the expansion coefficients are functions of $u$ and $x^A$, and their indices are raised and lowered by $\gamma_{AB}$ and its inverse $\gamma^{AB}$, respectively.
The covariant derivative $\mathscr D_A$ is the one compatible with $\gamma_{AB}$.
Among these coefficients, $m$ and $N^A$ are called the Bondi mass and angular momentum aspects as in general relativity, respectively \cite{Bondi:1962px,Sachs:1962wk}.
$c_{AB}$ is the shear tensor associated with the outgoing null geodesic congruence.
Because of the determinant condition \eqref{eq-det-gr}, one finds out that \cite{Flanagan:2015pxa}
\begin{gather}
    c_{AB}=\hat c_{AB}, \quad d_{AB}=\hat d_{AB}+\frac{1}{4}\gamma_{AB} c_{C}^D c^{C}_D,\\
    e_{AB}=\hat e_{AB}+\frac{1}{2}\gamma_{AB} c_{C}^D\hat d^{C}_D,
\end{gather}
with the hatted tenors being traceless with respect to $\gamma^{AB}$.

Since the volume $\epsilon_{abcd}$ is present in the equations of motion, its component form should be discussed.
So first, let $\hat\epsilon_{AB}$ be the volume element compatible with $\gamma_{AB}$, so then one has 
\begin{equation}
    \label{eq-2vol}
    \hat\epsilon_{\theta\phi}=\sqrt{\gamma}=\sin\theta,
\end{equation}
where $\gamma$ is the determinant of $\gamma_{AB}$.
Thus,
\begin{equation}
    \label{eq-4vol}
    \epsilon_{ur\theta\phi}=\sqrt{-g}=e^{2\beta}r^2\sqrt{\gamma}.
\end{equation}
So both $\epsilon_{abcd}$ and $\hat\epsilon_{AB}$ have simple asymptotic behaviors.

With these boundary conditions \eqref{eq-exp-beta} - \eqref{eq-thex} and \eqref{eq-4vol}, one can check that C-tensor has the following asymptotic behavior,
\begin{gather*}
    C_{uu}=\order{r^{-3}}, \quad C_{ur}=\order{r^{-4}}, \quad C_{rr}=\order{r^{-5}},\\ C_{uA}=\order{r^{-2}},\quad C_{rA}=\order{r^{-3}},\quad C_{AB}=\order{r^{-2}}.
\end{gather*}
Therefore, the boundary condition of $C_{ab}$ is consistent with the one in Ref.~\cite{Flanagan:2015pxa}, which justifies the use of Eqs.~\eqref{eq-exp-beta} - \eqref{eq-thex}.

\subsection{Asymptotic solutions}
\label{sec-asol}

Given the chosen asymptotic behaviors, one can solve the equations of motion  by directly substituting asymptotic expansions into the equations of motion to determine the relations among the expansion coefficients.
After some complicated algebraic manipulations, one finds that $\vartheta_0$ is constant, which can be set to zero due to the shift symmetry. 
Furthermore, one obtains the following results:
    \begin{gather}
        \beta_1=0,\\
        \beta_2=-\frac{1}{32} c_{AB} c^{AB}-\frac{\mathfrak b}{16\kappa}\vartheta_1^2,\\
        \beta_3=-\frac{\mb}{6\kappa}\vartheta_1\vartheta_2,\\
        \mathcal U_A=-\frac{1}{2}\mathscr D^B c_{AB},\\
        6\mathcal M+\frac{3}{16} c_{AB} c^{AB}+\mathscr D_AN^A+\frac{3}{4}\mathscr D_A c^{AB}\mathscr D^C c_{BC}\nonumber\\
        +\frac{3\mb}{8\kappa}\left( \vartheta_1^2+\mathscr D_A\vartheta_1\mathscr D^A\vartheta_1-\vartheta_1\mathscr D^2\vartheta_1 \right)=0,\label{eq-fl2.20}\\
        \hat d_{AB}=0,\label{eq-fl2.10}
    \end{gather}
where $\mathscr D^2=\mathscr D_A\mathscr D^A$.
One also obtains the following evolution equations,
    \begin{gather}
        \dot\vartheta_2=-\frac{1}{2}\mathscr D^2\vartheta_1,\label{eq-evo-th2}\\
        \dot m=\frac{1}{4}\mathscr D_A\mathscr D_BN^{AB}-\frac{1}{8}N_{AB}N^{AB}-\frac{\mb}{4\kappa}N^2,\label{eq-evo-m}\\
        \begin{split}
        \dot N_A=&\mathscr D_Am+\frac{1}{4}(\mathscr D_B\mathscr D_A\mathscr D_C c^{BC}-\mathscr D_B\mathscr D^B\mathscr D_C c_A^C)\\
        &+\frac{1}{4}\mathscr D_C(N^{BC} c_{AB})+\frac{1}{2} c_{AB}\mathscr D_CN^{BC}\\
        &+\frac{\mb}{8\kappa}\left( \vartheta_1\mathscr D_AN-3N\mathscr D_A\vartheta_1 \right),\label{eq-evo-n}
        \end{split}
    \end{gather}
where the dot means to take the derivative $\pd/\pd u$, and
\begin{equation}
    \label{eq-def-news}
    N_{AB}=\dot c_{AB}, \quad N=\dot\vartheta_1.
\end{equation}
The equations for $\hat e_{AB}$ and $\mathscr U^A$ are way more complicated and relegated into Appendix~\ref{sec-app}.

One can find out that $c_{AB}$ and $\vartheta_1$ have no evolution equations, and all of the above equations are written in terms of them and their derivatives. 
In fact, they represent the physical degrees of freedom of the theory as in Brans-Dicke theory \cite{Hou:2020tnd,Tahura:2020vsa,Hou:2020tnd}. 
Since $c_{AB}$ is a symmetric and traceless rank-2 tensor on the unit 2-sphere, there are three degrees of freedom in this theory.
Like in general relativity and Brans-Dicke theory \cite{Bondi:1962px,Sachs:1962wk,Hou:2020tnd,Tahura:2020vsa}, the vanishing of $N_{AB}$ and $N$ means the absence of gravitational waves.

From the above expressions, one also finds out that up to the orders considered in Eqs.~\eqref{eq-exp-beta} - \eqref{eq-thex} and ignoring equations in Appendix~\ref{sec-app}, the asymptotically flat spacetime at the null infinity looks just like the one in general relativity with a canonical scalar field $\sqrt{\mb}\vartheta$, which couples with the metric $g_{ab}$ minimally \cite{Flanagan:2015pxa}.
The nonminimal coupling between $\vartheta$ and $g_{ab}$ occurs in those equations in Appendix~\ref{sec-app}, and, of course, terms at even higher orders.
Because of the complicated forms of the higher order equations, none of them will be presented here.
Note the appearance of $\hat\epsilon_{AB}$ in Eqs.~\eqref{eq-u4}, \eqref{eq-eevo} and \eqref{eq-evo-u4}, which represents the parity violation effect.

In the end, for the purpose of reference, we present the remaining metric components,
    \begin{gather}
        \begin{split}
        g_{uu}=&-1+\frac{2m}{r}-\frac{1}{r^2}\left[ \frac{\mathscr D_AN^A}{3}\right.\\
        &\left.+\frac{\mb}{8\kappa}(\mathscr D_A\vartheta_1\mathscr D^A\vartheta_1-\vartheta_1\mathscr D^2\vartheta_1) \right]+\order{\frac{1}{r^3}},
        \end{split}\\
        \begin{split}
        g_{ur}=&-1+\frac{1}{r^2}\left( \frac{c_{AB}c^{AB}}{16}+\frac{\mb}{8\kappa}\vartheta_1^2 \right)\\
        &+\frac{\mb}{3\kappa r^3}\vartheta_1\vartheta_2+\order{\frac{1}{r^4}},
        \end{split}\\
        \begin{split}
        g_{uA}=&\frac{\mathscr D_Bc_{A}^B}{2}+\frac{1}{r}\left( \frac{2}{3}N_A-\frac{1}{8}c_{BC}\mathscr D_Ac^{BC} \right)\\
           &-\frac{1}{r^2}\left[\mathscr U_A-\frac{2}{3}c_{AB}N^B +\frac{1}{8}\left(c_C^Dc^C_D\mathscr D_Bc_A^B\right.\right.\\
           &\left.\left.+\frac{1}{2}c_A^Bc_C^D\mathscr D_Bc_D^C\right)  \right]+\order{\frac{1}{r^3}}
        \end{split}
    \end{gather}
In these expressions, the parity violating terms explicitly appear at the $1/r^2$ order in $g_{uA}$ due to the presence of $\mathscr U_A$ which is given by Eq.~\eqref{eq-u4} and contains $\hat\epsilon_{AB}$.

\subsection{Asymptotic symmetries}
\label{sec-bms}

Due to the similarity mentioned previously, many conclusions valid in general relativity also hold in Chern-Simons gravity.
In particular, the asymptotic symmetries are actually the same. 
This is due to the fact that the asymptotic symmetries are diffeomorphisms that preserve the boundary conditions Eqs.~\eqref{eq-guu-bdy} - \eqref{eq-h-bdy} and the determinant condition \eqref{eq-det-gr}, and these conditions take exactly the same forms as in general relativity \cite{Barnich:2010eb}.
So, let the vector field $\xi^a$ generate an infinitesimal asymptotic symmetry such that
    \begin{gather}
        \lie_\xi g_{rr}=\lie_\xi g_{rA}=0,\label{eq-bms-inf-0}\\
        g^{AB}\lie_\xi g_{AB}=0,\label{eq-bms-inf-1}\\
        \lie_\xi g_{ur}=\order{r^{-1}},\quad \lie_\xi g_{uA}=\order{1},\\ \lie_\xi g_{AB}=\order{r},\label{eq-bms-inf-2}\\ 
        \lie_\xi g_{uu}=\order{r^{-1}},\label{eq-bms-inf-3}\\
        \lie_\xi\vartheta=\order{r^{-1}}.\label{eq-bms-inf-s}
    \end{gather}
Like in general relativity and Brans-Dicke theory \cite{Barnich:2010eb,Hou:2020tnd}, $\xi^a$ has the following components
    \begin{gather}
        \xi^u=f(u,x^A),\label{eq-xi-u}\\ 
        \xi^A=Y^A(u,x^A)-(\mathscr D_Bf)\int_r^\infty e^{2\beta}g^{AB}\ud r',\label{eq-xi-a}\\
        \xi^r=\frac{r}{2}(U^A\mathscr D_Af-\mathscr D_A\xi^A),\label{eq-xi-r}
    \end{gather}
where $f$ and $Y^A$ are arbitrary integration functions independent of $r$.
These components are obtained by evaluating Eqs.~\eqref{eq-bms-inf-0} and \eqref{eq-bms-inf-1}.
Using the asymptotic expansions of the metric functions and the scalar field, one knows that 
    \begin{gather}
        \begin{split}
        \xi^A=&Y^A-\frac{\mathscr D^Af}{r}+\frac{c^{AB}\mathscr D_Bf}{2r^2}\\
        &-\frac{1}{r^3}\left( \frac{c_{B}^Cc^{B}_C}{16}-\frac{\mb}{24\kappa}\vartheta_1^2 \right)\mathscr D^Af+\order{\frac{1}{r^4}},
        \end{split}\\
        \begin{split}
        \xi^r=&-\frac{r}{2}\psi+\frac{1}{2}\mathscr D^2f-\frac{1}{2r}\bigg[ (\mathscr D_Af)\mathscr D_Bc^{AB}\\
        &+\frac{1}{2}c^{AB}\mathscr D_A\mathscr D_Bf \bigg]+\order{\frac{1}{r^2}}.
        \end{split}
    \end{gather}
Again, the parity violating terms are of the higher orders.
The remaining conditions imply that $Y^A$ is actually independent of $u$ and is a conformal Killing vector field for $\gamma_{AB}$, i.e.,
\begin{equation}\label{eq-pro-conf}
        \lie_Y\gamma_{AB}=\psi\gamma_{AB}, \quad \psi=\mathscr D_AY^A.
\end{equation}
Finally, one also finds out that there exists an arbitrary function $\alpha(x^A)$ independent of $u$ such that
\begin{equation}
    \label{eq-f-a}
    f=\alpha+\frac{u}{2}\psi.
\end{equation}
These two equations take the same forms as in general relativity and Brans-Dicke theory.
Therefore, one calls the diffeomorphisms generated by $\alpha$ supertranslations, generalizing the familiar translations in Minkowski spacetime.
In fact, if $\alpha$ is a linear combination of $l=0,1$ spherical harmonics, it generates the usual space and time translation.
A generic supertranslation is given by a linear combination of all spherical harmonics.
The transformations generated by the global conformal Killing vector fields $Y^A$ form a group isomorphic to the Lorentz group.
One can rewrite $Y^A$ in the following way,
\begin{equation}
    \label{eq-decy}
    Y^A=\mathscr D^A\chi+\hat\epsilon^{AB}\mathscr D_B\sigma,
\end{equation}
with $\chi$ and $\sigma$ linear combinations of $\ell=1$ spherical harmonics, and then $\chi$ generates boost and $\sigma$ generates (spatial) rotation \cite{Flanagan:2015pxa}.
The semidirect sum of the supertranslation group and the Lorentz group is the celebrated BMS group \cite{Bondi:1962px,Sachs:1962zza}.
One can also choose to work in complex stereographic coordinates $(\zeta,\bar \zeta)$ where $\zeta=e^{i\phi}\cot(\theta/2)$ and bar means to take the complex conjugation. 
Then, Eq.~\eqref{eq-pro-conf} becomes $\pd_\zeta Y^{\bar\zeta}=\pd_{\bar\zeta}Y^{\zeta}=0$, so $Y^\zeta=Y^\zeta(\zeta)$ and $Y^{\bar\zeta}=Y^{\bar\zeta}(\bar\zeta)$.
One usually chooses the following bases for $Y^A$ \cite{Blumenhagen:2009zz},
\begin{equation}
    \label{eq-def-by}
    l_n=-\zeta^{n+1}\pd_\zeta,\quad \bar l_n=-\bar\zeta^{n+1}\pd_{\bar\zeta},
\end{equation}
where $n$ are integers, in general.
A global conformal Killing vector field is a linear combination of $l_{-1},\,l_0,\,l_1$ and $\bar l_{-1},\,\bar l_0,\,\bar l_1$.
If one allows all conformal Killing vectors, i.e., $n$ takes all integral values, the Lorentz algebra is extended to the Virasoro algebra, and the resultant symmetry group is called the extended BMS group \cite{Flanagan:2015pxa}.
Superboosts and superrotations, generalizations of boosts and rotations, are just the linear combinations of $l_n$ and $\bar l_n$ for any integers $n$.
For completeness, the supertranslation generator $\alpha$ is a linear combination of $2\zeta^n\bar\zeta^{\bar n}/(1+\zeta\bar\zeta)$ where $n,\,\bar n$ are both integers \cite{Barnich:2010eb}.
When $n,\,\bar n=0,1$, one obtains the generators for the usual translations.

Therefore, the extended BMS transformation takes the Bondi-Sachs coordinate system into a new Bondi-Sachs coordinate system.
A particular Bondi-Sachs coordinate system is called a BMS frame.
Under the asymptotic symmetry transformation generated by $\xi^a$, a BMS frame is transformed to a new one, and the metric $g_{ab}$ and the scalar field $\vartheta$ transform according to $\delta_\xi g_{ab}=\lie_\xi g_{ab}$ and $\delta_\xi\vartheta=\lie_\xi\vartheta$, respectively.
Therefore, one obtains following transformation laws:
    \begin{gather}
    \delta_\xi\vartheta_1=fN+\frac{\psi}{2}\vartheta_1+\lie_Y\vartheta_1,    \label{eq-bms-phi1}\\
\begin{split}
        \delta_\xi c_{AB}=&fN_{AB}-2\mathscr D_A\mathscr D_Bf+\gamma_{AB}\mathscr D^2f\\
        &+\lie_Y c_{AB}-\frac{\psi}{2}c_{AB},\label{eq-bms-c}
\end{split}\\
        \begin{split}
        \delta_\xi m=&f\dot m+\lie_Ym+\frac{3}{2}\psi m+\frac{1}{8}c^{AB}\mathscr D_A\mathscr D_B\psi\\&+\frac{1}{4}N^{AB}\mathscr D_A\mathscr D_Bf+\frac{1}{2}(\mathscr D_Af)\mathscr D_BN^{AB},
        \end{split}\\
         \begin{split}
            \label{eq-bms-n}
        \delta_\xi N_A=&f\dot N_A+\lie_Y N_A+\psi N_A+3m\mathscr D_Af\\
           &+\frac{3}{4}(\mathscr D_A\mathscr D_Cc^C_B-\mathscr D_B\mathscr D_Cc^C_A)\mathscr D^Bf\\
           &+\frac{3}{4}c_{AC}N^{BC}\mathscr D_Bf+\frac{\mb}{8\kappa}\bigg( \vartheta_1N\mathscr D_Af\\
           &-\frac{1}{2}\vartheta_1^2\mathscr D_A\psi \bigg),
        \end{split}
    \end{gather}
where the symbol $\lie_Y$ is to take the Lie derivative on the unit 2-sphere.
Then, the news tensor $N_{AB}$ and the scalar $N$ transform according to 
    \begin{gather}
        \delta_\xi N_{AB}=f\dot N_{AB}+\lie_YN_{AB},\\ 
        \delta_\xi N=f\dot N+\psi N+\lie_YN.
    \end{gather}
From the last two equations, one knows that if there are no gravitational waves in a certain region of the future null infinity in one BMS frame, i.e., $N_{AB}=0=N$, these two quantities still vanish in a different BMS frame.

\section{Memory effects}
\label{sec-mm}

Memory effects generally refer to the permanent change in the relative distance between two test particles after the passage of gravitational waves \cite{Zeldovich:1974gvh,Braginsky:1986ia,Christodoulou1991,Thorne:1992sdb}. 
This particular phenomenon is also called the displacement memory, as there were new memory effects discovered more recently. 
Among them, spin memory effect and CM memory effect  \cite{Pasterski:2015tva,Nichols:2018qac} will also be considered in this work.
In the following, these effects will be presented by considering the relative motion between test particles due to the presence of gravitational waves.

\subsection{Geodesic deviation}
\label{sec-gde}

Since in Sec.~\ref{sec-cs}, one assumes there is no direct interaction between $\vartheta$ and the ordinary matter fields, the relative acceleration of two test particles is simply due to the spacetime curvature \cite{Wald:1984rg},
\begin{equation}
    \label{eq-gde}
    T^c\nabla_c(T^b\nabla_bS^a)=-R_{cbd}{}^aT^cS^bT^d,
\end{equation}
where $T^a=(\ud/\ud\tau)$ is the four-velocity of a freely falling test particle with $\tau$ being the proper time, and where $S^a$ is the deviation vector between adjacent test particles.
This is just the geodesic deviation equation, and it is used to detect gravitational waves by interferometers \cite{Misner:1974qy}.
Applying this equation to the test particles near the null infinity of the spacetime considered in the previous section, one obtains the following relative acceleration \cite{Hou:2020xme}:
\begin{equation}
    \label{eq-gde-c}
    \ddot S^{\hat A}\approx-R_{u\hat Bu}{}^{\hat A}S^{\hat B}=\frac{\ddot c_{\hat B}^{\hat A}}{2r}S^{\hat B}+\order{\frac{1}{r^2}}.
\end{equation}
Here, in writing down this equation, one actually sets up an orthonormal tetrad basis which contains basic vectors
\begin{gather*}
    T^a=(\pd_u)^a+\order{1/r},\\ (e_{\hat r})^a=-(\pd_u)^a+(\pd_r)^a+\order{1/r},\\
    (e_{\hat \theta})^a=r^{-1}(\pd/\pd \theta)^a+\order{1/r^2},\\ (e_{\hat\phi})^a=(r\sin\theta)^{-1}(\pd/\pd\phi)^a+\order{1/r^2},
\end{gather*}
so that the indices $\hat A,\,\hat B=\hat \theta,\,\hat \phi$.
Also, the proper time $\tau$ approaches the retarded $u$ as the test particles are close to the null infinity \cite{Strominger:2014pwa}.
From Eq.~\eqref{eq-gde-c}, one can find out that although there are three gravitational degrees of freedom in Chern-Simons gravity, the interferometer can detect only two of them, i.e., the plus and cross polarizations. 
This is drastically different from the situations in other modified gravities, where each gravitational degree of freedom would excite its own polarizations that can be detected by interferometers, pulsar timing arrays,  and the Gaia mission \cite{Gong:2017bru,Hou:2018djz,Gong:2018cgj,Gong:2018vbo,Gong:2018ybk}. 
For example, in scalar-tensor theories, there exists an extra polarization named breathing mode if the scalar field is massless \cite{Will:2014kxa,Liang:2017ahj,Hou:2017bqj}.

Using the geodesic deviation equation, one can introduce memory effects.
Let us assume that there is no gravitational wave before $u_0$ and after $u_f$, during which $N_{\hat A\hat B}=0$ and $N=0$.
These spacetime regions are said to be nonradiative.
Then, integrating this equation twice, one obtains,
    \begin{gather}
        \dot S^{\hat A}(u)\approx\dot S^{\hat A}_0+\frac{1}{2r}\int_{u_0}^{u}\ud u'\ddot c_{\hat B}^{\hat A}S^{\hat B}(u'),\label{eq-vf}\\
        \begin{split}
        S^{\hat A}(u)\approx&S^{\hat A}_0+(u-u_0)\dot S^{\hat A}_0\\
        &+\frac{1}{2r}\int_{u_0}^{u}\ud u'\int_{u_0}^{u'}\ud u''\ddot c_{\hat B}^{\hat A}S^{\hat B}(u''),\label{eq-sf}
        \end{split}
    \end{gather}
where $\dot S_0^{\hat A}$ and $S_0^{\hat A}$ are the initial relative velocity and the initial relative displacement, respectively.
Substituting Eq.~\eqref{eq-sf} back into itself and Eq.~\eqref{eq-vf}, one finds the following total changes at the time $u>u_f$,
    \begin{gather}
        \Delta\dot S_{\hat A}\approx-\frac{\Delta c_{\hat A\hat B}}{2r}\dot S_0^{\hat B},\label{eq-def-vk}\\
        \begin{split}
        \Delta S_{\hat A}&\approx\dot S_{\hat A}^0\Delta u+\frac{\Delta c_{\hat A\hat B}}{2r}S^{\hat B}_0\\
        &+\frac{1}{r}\left[\frac{c_{\hat A\hat B}(u_f)+c_{\hat A\hat B}(u_0)}{2}\Delta u-\Delta\mathcal C_{\hat A\hat B}\right]\dot S_0^{\hat B} ,\label{eq-def-3mm}
        \end{split}
    \end{gather}
where  $\Delta u=u_f-u_0$, $\Delta c_{\hat A\hat B}=c_{\hat A\hat B}(u_f)-c_{\hat A\hat B}(u_0)$, and 
\begin{equation}
    \label{eq-def-sc}
    \Delta\mathcal C_{\hat A\hat B}=\int_{u_0}^{u_f}c_{\hat A\hat B}(u)\ud u.
\end{equation}
Equation~\eqref{eq-def-3mm} takes a different form from (4.17) in \cite{Tahura:2020vsa} and (2.23) in \cite{Seraj:2021qja} with the effect of the Brans-Dicke scalar field ignored, but they are all equivalent.

From the above equations, one realizes that as long as there exists the relative velocity and displacement initially, the final relative velocity and displacement will change permanently even after the gravitational wave disappears.
This phenomenon is the memory effect. 
More specifically, Eq.~\eqref{eq-def-vk} describes the \emph{velocity kick memory} \footnote{Note the difference between the velocity \emph{kick} memory named by Ref.~\cite{Seraj:2021qja} and the velocity memory defined in \cite{Zhang:2017rno,Zhang:2017geq,Zhang:2018srn,Compere:2018ylh,Mao:2019sph}. The velocity kick memory is present only if the initial relative velocity $S^A_0$ is nonzero, while the existence of the velocity memory has nothing to do with the initial relative velocity. In fact, even if the initial relative velocity is zero, the velocity memory could also exist.}, and Eq.~\eqref{eq-def-3mm} is the displacement memory.
In particular, the second term on the right-hand side of Eq.~\eqref{eq-def-3mm} is the \emph{leading} displacement memory effect, the first memory effect discovered a long time ago \cite{Zeldovich:1974gvh,Braginsky:1986ia,Christodoulou1991,Thorne:1992sdb}.
Both the velocity kick and the leading displacement memories are due to the change in $c_{\hat A\hat B}$.
The second line of Eq.~\eqref{eq-def-3mm} is the \emph{subleading} displacement memory, which also depends on $\Delta\mathcal C_{\hat A\hat B}$, the time integral of $c_{\hat A\hat B}$.
In fact, the electric and magnetic parts of $\Delta\mathcal C_{\hat A\hat B}$ are related to the spin and CM memory effects, as discussed below \cite{Flanagan:2015pxa,Hou:2020wbo,Tahura:2020vsa}.
The subleading displacement memory was also studied in Refs.~\cite{Compere:2019odm,Mao:2020vgh} within the framework of general relativity.

As revealed in Eq.~\eqref{eq-gde-c}, the interferometer cannot detect the scalar gravitational wave polarization caused by $\vartheta$, so it cannot measure the memory due to $\vartheta$ even if it  exists.
In fact, one cannot use pulsar timing arrays or Gaia mission to detect the scalar memories, either.
In addition, the proper description of the scalar memory effect might rely on the dual formalism of scalar fields \cite{Campiglia:2017dpg,Campiglia:2018see,Yoshida:2019dxu,Seraj:2021qja}, which is beyond the scope of this work.
So here, the scalar memory effect will not be discussed.

\subsection{Vacuum transitions}
\label{sec-vt}

Using the transformation law \eqref{eq-bms-c}, one can relate the velocity kick and the leading displacement memory effects to vacuum transitions of the gravitational system at the null infinity \cite{Strominger:2014pwa,Hou:2020tnd}, as they both depend on $\Delta c_{AB}$.
The definition of the vacuum state in gravitational systems is not trivial.
Although it is easy to understand that in a vacuum state, $N=0$, i.e., there is no scalar gravitational wave, it is a little more involved to determine the correct conditions for $c_{AB}$.
Here, one would impose the same conditions as in general relativity \cite{Strominger:2014pwa,Hou:2020tnd}. 
So one first needs the following Newman-Penrose tetrad basis $\{l^a,n^a,m^a,\bar m^a\}$ \cite{Newman:1961qr}:
\begin{gather}
    l^a=(\pd_r)^a+\order{r^{-1}},\\ n^a=-(\pd_u)^a+\frac{1}{2}(\pd_r)^a+\order{r^{-1}},\\
    m^a=\frac{1}{\sqrt{2}r}\left[ (\pd_\theta)^a-i\csc\theta (\pd_\phi)^a \right]+\order{\frac{1}{r^2}},
\end{gather}
and $\bar m^a$ is the complex conjugate of $m^a$.
Then one requires that the following Newman-Penrose variables vanish at the leading order:
    \begin{gather}
        \Psi_4=C_{abcd}n^a\bar m^bn^c\bar m^d=-\frac{r}{2}\pd_uN_{AB}\bar m^{A}\bar m^{B}+\cdots,\\ 
        \Psi_3=C_{abcd}\bar m^an^bl^cn^d=-\frac{1}{2r}\bar m^A\mathscr D_BN^B_A+\cdots,\\
        \begin{split}\label{eq-impsi2}
        \Im\Psi_2=&\Im(C_{abcd}\bar m^an^bl^cm^d)\\ 
        =&\frac{1}{i8r}(N_A^C\hat c_{BC}-\mathscr D_A\mathscr D_C\hat c^C_B+\mathscr D_B\mathscr D_C\hat c^C_A)\times\\
        &(\bar m^Am^B-m^A\bar m^B)+\cdots,
        \end{split}
    \end{gather}
where dots represent higher order terms, and $\Im$ is to take the imaginary part.
Therefore, one knows that $N_{AB}=0$ and 
\begin{equation}
    \label{eq-cvac}
    \tilde c_{AB}=\left(\mathscr D_A\mathscr D_B-\frac{1}{2}\gamma_{AB}\mathscr D^2\right)\Phi(x^C),
\end{equation}
for some arbitrary function $\Phi$ on the unit 2-sphere.
Here, a tilde means to evaluate $c_{AB}$ in the vacuum state.

To find the relation between the memory effect and the vacuum transition, it is sufficient to rewrite Eq.~\eqref{eq-bms-c} in the nonradiative region, and in particular, to consider the supertranslation transformation generated by $\alpha$ as
\begin{equation}
    \label{eq-st-c}
    \delta_\alpha \tilde c_{AB}=-2\mathscr D_A\mathscr D_B\alpha+\gamma_{AB}\mathscr D^2\alpha.
\end{equation}
This implies that the transformed $c_{AB}$ still describes a vacuum state $\tilde c'_{AB}$ with $\Phi'=\Phi-2\alpha$.
Just like in general relativity and Brans-Dicke theory \cite{Strominger:2014pwa,Hou:2020tnd}, there are also infinitely many degenerate vacuum states that can be transformed to each other via supertranslations.
The vacuum transition causes the change in $c_{AB}$, 
\begin{equation}
    \label{eq-cc}
    \Delta c_{AB}=\left(\mathscr D_A\mathscr D_B-\frac{1}{2}\gamma_{AB}\mathscr D^2\right)\Delta\Phi=\delta_\alpha\tilde c_{AB},
\end{equation}
which explains the velocity kick and the leading order displacement memories.

\subsection{Constraints on memory effects}
\label{sec-cmm}

From the above discussion, one knows that memory effects are related to $\Delta c_{AB}$ and $\mathcal C_{AB}$.
These two quantities are constrained by conservation laws associated with the extended BMS symmetries \cite{Flanagan:2015pxa,Hou:2020wbo,Tahura:2020vsa,Hou:2020xme}.
To determine the constraints requires us to calculate the conserved charges and the fluxes using certain formalisms such as the one in Ref.~\cite{Wald:1999wa}, which is very involved and will be done in a subsequent paper.
In fact, there is a second method to constraining memory effects by properly integrating the evolution equations \eqref{eq-evo-m} and \eqref{eq-evo-n} \cite{Hou:2020tnd}.

First, substituting $c_{AB}=(\mathscr D_A\mathscr D_B-\frac{1}{2}\gamma_{AB}\mathscr D^2)\Phi+\hat\epsilon_{C(A}\mathscr D_{B)}\mathscr D^C\Psi$ into the first $N^{AB}$ in Eq.~\eqref{eq-evo-m}, multiplying both sides by an arbitrary supertranslation generator $\alpha$, and integrating the resulting equation over the null infinity, one obtains 
\begin{equation}
    \label{eq-cons-phi}
    \begin{split}
    &\oint\ud^2\Omega\alpha\mathscr D^2(\mathscr D^2+2)\Delta\Phi\\
    =&\oint\ud^2\Omega\left[8\alpha \Delta m+\int_{u_0}^{u_f}\ud u\alpha\left( N_{AB}N^{AB}+\frac{2\mb}{\kappa}N^2 \right)\right],
    \end{split}
\end{equation} 
where $\ud^2\Omega=\sin\theta\ud\theta\ud\phi$.
This gives the constraint on the velocity kick and leading displacement memory effects.
In fact, one could guess from the form of the equation that the terms in the square brackets are proportional to the energy density of the tensor and the scalar gravitational waves. 
In literature, these terms are said to cause the null memory, and the one with $\Delta m$ causes the ordinary memory \cite{Bieri:2013ada}.

Second, to obtain the constraints on the subleading displacement memory, or on the spin and CM memories, one might want to modify Eq.~\eqref{eq-evo-n} in the following way:
\begin{equation}
    \label{eq-mevo-n}
    \begin{split}
        \pd_u\hat N_A=&\mathscr D_Am+\frac{1}{4}\hat\epsilon_{AB}\mathscr D^B\eta+\frac{1}{4}\mathscr D_A(c_C^DN_D^C)\\
        &-\frac{1}{4}N^C_D\mathscr D_Ac_C^D-\frac{1}{4}\hat\epsilon_{AB}\mathscr D^B\rho\\
        &+\frac{\mb}{8\kappa}\left( \vartheta_1\mathscr D_AN-3N\mathscr D_A\vartheta_1 \right),
    \end{split}
\end{equation}
with $\hat N_A=N_A-\frac{3}{32}\mathscr D_A(c_B^Cc^B_C)-\frac{1}{4}c_{AB}\mathscr D_Cc^{BC}$ and $\rho=\hat\epsilon^{AB}N_A^Cc_{BC}$.
Now, the equation is similar to Eq.~(4.49) in Ref.~\cite{Barnich:2010eb} in general relativity, neglecting the last line here and setting $l=0$ and $\bar R=2$ there.
Then it is easy to determine the constraint on the spin memory which is measured by \cite{Flanagan:2015pxa}
\begin{equation}
    \label{eq-def-sm}
    \Delta\mathcal S=\int_{u_0}^{u_f}\ud u\Psi.
\end{equation}
That is, one can contract both sides by $\hat\epsilon^{AB}\mathscr D_B\sigma$, which is the magnetic parity of $Y^A$ [refer to Eq.~\eqref{eq-decy}],
then perform the integral over the null infinity to arrive at
\begin{equation}
    \label{eq-spin-cs}
    \begin{split}
    &\oint\ud^2\Omega\sigma\mathscr D^2\mathscr D^2(\mathscr D^2+2)\Delta\mathcal S\\
    =&-\oint\ud^2\Omega\bigg[\sigma\hat\epsilon^{AB}\mathscr D_A\Delta\hat N_B\\
    &+\int_{u_0}^{u_f}\ud u\sigma\left(\frac{1}{4}\mathscr D^2\rho-\hat\epsilon^{AB}\mathscr D_AJ_B  \right)\bigg],
    \end{split}
\end{equation}
where $\Delta\hat N_A=\hat N_A(u_f)-\hat N_A(u_0)$, and
\begin{equation}
    \label{eq-and}
    J_A=\frac{1}{4}N_D^C\mathscr D_Ac_C^D+\frac{\mb}{2\kappa}N\mathscr D_A\vartheta_1.
\end{equation}
Formally, $J_A$ is proportional to the angular momentum flux density of the tensor and the scalar gravitational waves.
In the end, one tries to obtain the constraint on the CM memory \cite{Nichols:2018qac}.
This is a bit more complicated. 
One should notice that the operator $\mathscr D^2(\mathscr D^2+2)$ on the left-hand side of Eq.~\eqref{eq-cons-phi} is linear, so one can split $\Phi$ into two parts, $\Phi=\Phi_o+\Phi_n$ so that $\Delta\Phi_o$ is caused by $\Delta m$, and $\Delta\Phi_n$ is caused by the remaining parts on the right-hand side of Eq.~\eqref{eq-cons-phi}.
Since Eq.~\eqref{eq-cons-phi} comes from the evolution equation \eqref{eq-evo-m}, one may identify the following relation:
\begin{equation}
    \label{eq-evo-m-o}
    \dot m=\frac{1}{8}\mathscr D^2(\mathscr D^2+2)\dot\Phi_o,
\end{equation}
because $\mathscr D_A\mathscr D_BN^{AB}=\mathscr D^2(\mathscr D^2+2)\dot\Phi/2$.
Now, choose $Y^A=\mathscr D^A\chi$, i.e., the electric parity part. 
The infinitesimal extended BMS transformation generated by this $Y^A$ has $f=u\psi/2=u\mathscr D^2\chi/2$, according to Eq.~\eqref{eq-f-a}.
Multiplying Eq.~\eqref{eq-evo-m-o} by $f$, contracting Eq.~\eqref{eq-mevo-n} by $\mathscr D^A\chi$, and combining the results properly, one obtains
\begin{equation}
    \label{eq-cm-cs}
    \begin{split}
    &\oint\ud^2\Omega\chi\mathscr D^2\mathscr D^2(\mathscr D^2+2)\Delta\mathcal K\\
    =&\oint\ud^2\Omega\chi \Bigg\{8\Delta\left[u\mathscr D^2m-\mathscr D^A\hat N_A\right]\\
        +&\int_{u_0}^{u_f}\ud u\left[\mathscr D^2\left(2c_A^BN_B^A+\frac{\mb}{\kappa}\vartheta_1N\right)-8\mathscr D^AJ_A\right]\Bigg\},
    \end{split}
\end{equation}
where the CM memory is quantified by 
\begin{equation}
    \label{eq-def-cm}
    \Delta\mathcal K=\int_{u_0}^{u_f}u\dot\Phi_o\ud u.
\end{equation}
Note that in Eqs.~\eqref{eq-spin-cs} and \eqref{eq-cm-cs}, both $\sigma$ and $\chi$ are arbitrary functions on the unit 2-sphere, not just linear combinations of $l=1$ spherical harmonics.
As a final remark, one notices that 
\begin{equation}
    \label{eq-dec-mc}
    \begin{split}
   \Delta \mathcal C_{AB}
        =&\left(\mathscr D_A\mathscr D_B-\frac{1}{2}\gamma_{AB}\mathscr D^2\right)\left[ u_f\Delta\Phi+\Phi(u_0)\Delta u\right.\\
        &\left.-\Delta\mathcal L-\Delta\mathcal K \right]+\hat\epsilon_{C(A}\mathscr D_{B)}\mathscr D^C\Delta\mathcal S,
    \end{split}
\end{equation}
where $\Delta\mathcal L=\displaystyle\int_{u_0}^{u_f}u\dot\Phi_n\ud u$. 
By putting Eq.~\eqref{eq-evo-m-o} into Eq.~\eqref{eq-evo-m},  multiplying the resultant equation by $u\omega(x^A)$ with $\omega$ being any function on the 2-sphere, and performing the integral, one gets 
\begin{equation}
    \label{eq-uni}
    \begin{split}
    &\oint\ud^2\Omega\omega\mathscr D^2(\mathscr D^2+2)\Delta\mathcal L\\
    =&\int_{u_0}^{u_f}\ud u\oint\ud^2\Omega\omega\left( N_A^BN_B^A+\frac{2\mb}{\kappa}N^2 \right).
    \end{split}
\end{equation}
Therefore, in a certain sense, the constraints on the spin and the CM memories also give the constraint on the subleading displacement memory.

To summarize, here, one properly integrates the evolution equations \eqref{eq-evo-m} and \eqref{eq-evo-n} multiplied by generators of the extended BMS transformations, then the constraints on various memory effects are determined. 
Usually, these constraints are expressed as fluxes and  charges associated with the extended BMS symmetry \cite{Compere:2019gft,Hou:2020wbo,Tahura:2020vsa,Hou:2020xme}.
So in principle, one can identify those charges and fluxes in the above constraint equations.
However, we will not do that here. 
Instead, the conserved charges and fluxes will be computed in a future work.

\section{Conclusion}
\label{sec-con}

This work discusses the asymptotically flat spacetime using Bondi-Sachs formalism and reveals memory effects predicted by the dynamical Chern-Simons gravity.
Like in general relativity and Brans-Dicke theory, the tensor gravitational degrees of freedom induce exactly the same kinds of memory effects.
That is, there are displacement, spin and CM memories.
The asymptotic symmetries of the spacetime are also the extended BMS symmetries, and they are related to these memories just like what happens in general relativity and Brans-Dicke theory.
So the displacement memory is related to the supertranslation transformation and the vacuum transition can be used to explain this effect.
It is constrained by Eq.~\eqref{eq-cons-phi}, where there are terms proportional to the energy densities of the tensor and scalar radiation, one of the conserved charges associated with supertranslations.
The spin memory is constrained by Eq.~\eqref{eq-spin-cs} and the CM memory by Eq.~\eqref{eq-cm-cs}.
Both equations contain derivatives of $J_A$ which is proportional to the angular momentum density and associated with the superboosts and superrotations in the extended BMS group.

Although there is one more gravitational degree of freedom  -- the Chern-Simons scalar field -- it does not excite memory effects that can be detected by interferometers, pulsar timing arrays, or the Gaia mission, due to the nonminimal interaction between it and the metric, and the absence of the direct coupling with the ordinary matter fields.
A similar situation would happen in more general parity violating theories, whose additional corrections are also of higher orders \cite{Qiao:2019wsh,Zhao:2019xmm,Kamada:2021kxi}.
However, it should have its own memory effects, the analysis of which should rely on some proper dual formalism that will be proposed in the future.

Besides Brans-Dicke theory and Chern-Simons gravity, there are more interesting modified theories of gravity, such as Einstein-\ae ther theory \cite{Jacobson:2000xp,Jacobson:2004ts}, Ho\v rava-Lifshitz gravity \cite{Horava:2009uw}, and so on.
The local Lorentz invariance is broken in these theories, and gravitational waves might have superluminal speeds \cite{Gong:2018cgj,Gong:2018vbo}.
Moreover, each degree of freedom excites its polarization that might be detected.
Whether there exist memories in these theories is to be answered.
Hopefully, the memory effect can be a new tool to tell the nature of gravity.

\begin{acknowledgements}
This work was supported by the National Natural Science Foundation of China under Grants No.~11633001, No.~11673008, No.~11922303, and No.~11920101003 and the Strategic Priority Research Program of the Chinese Academy of Sciences, Grant No. XDB23000000.
Tao Zhu is supported in part by the National Key Research and Development Program of China Grant No.~2020YFC2201503, the Zhejiang Provincial Natural Science Foundation of China under Grants No.~LR21A050001 and No.~LY20A050002.
\end{acknowledgements}

\appendix

\section{Some expansion coefficients}
\label{sec-app}

In this appendix, we will write down the equations for the remaining expansion coefficients of the metric functions. 
First, one has the equation for $\mathscr U_A$ appearing in Eq.~\eqref{eq-user} \cite{Godazgar:2018vmm},
\begin{equation}
    \label{eq-u4}
    \begin{split}
    \mathscr U_A=&\frac{1}{2}c_{AB}N^B+\frac{3}{4}\mathscr D_B\hat e_A^B-\frac{7}{64}\mathscr D_B(c_A^Bc_{C}^{D}c^{C}_{D})\\
    &-\frac{c_{C}^{D}c^{C}_{D}}{16}\mathscr D_Bc_A^B+\frac{\mb}{6\kappa}\left( \frac{3\vartheta_1}{16}\mathscr D_Bc_A^B+\vartheta_1\mathscr D_A\vartheta_2\right.\\
    &\left.-\frac{1}{2}\vartheta_2\mathscr D_A\vartheta_1 \right)+\frac{\ma}{4\kappa}\big[ \hat\epsilon_A{}^B\mathscr D_C(\vartheta_1N_B^C)\\
    &-\hat\epsilon_A{}^BN\mathscr D_Cc_B^C-\hat\epsilon^{BC}\pd_u(\vartheta_1\mathscr D_Bc_{AC})\\
    &+\hat\epsilon^{BC}N_{AC}\mathscr D_B\vartheta_1 \big].
    \end{split}
\end{equation}
The evolution equation for $\hat e_{AB}$ is \cite{Nichols:2018qac,Godazgar:2018vmm}
\begin{equation}\label{eq-eevo}
    \begin{split}
        \dot{\hat e}_{AB}=&\frac{1}{2}m c_{AB}+\frac{1}{4} c_{AB} c_{CD}N^{CD}+\frac{1}{6}(2\mathscr D_{(A}N_{B)}\\
        &-\gamma_{AB}\mathscr D_CN^C)+\frac{\mathfrak a}{2\kappa}\vartheta_1\hat\epsilon_{(A}{}^C\dot N_{B)C}\\
        &-\frac{1}{8}\left[\mathscr D_A\mathscr D_B-\frac{1}{2}\gamma_{AB}\mathscr D^2\right]( c_{CD} c^{CD})\\
        &-\frac{1}{8}\left( \mathscr D_A c_C^D\mathscr D_B c^C_D-\frac{1}{2}\gamma_{AB}\mathscr D_C c_{D}^{E}\mathscr D^C c^{D}_{E} \right)\\
        &-\frac{3}{8}\mathscr D_C(2  c^{CD}\mathscr D_{(A} c_{B)D} -\gamma_{AB} c^{CD}\mathscr D_E c^E_D)\\
        &-\frac{1}{4}( c_{C(A}\mathscr D^C\mathscr D^D c_{B)D}- c^{CD}\mathscr D_D\mathscr D_{(A} c_{B)C})\\
        &-\frac{3}{8} c_{AB}\mathscr D_C\mathscr D_D c^{CD}+\frac{1}{4}\mathscr D_C\mathscr D_D( c_{AB} c^{CD})\\
        &-\frac{1}{4}(\mathscr D_C c_A^C\mathscr D_D c_B^D+\mathscr D_C c_A^D\mathscr D_D c_B^C\\
        &-\mathscr D_C c_A^D\mathscr D^C c_{BD})-\frac{\mathfrak b}{8\kappa}\bigg[ \mathscr D_A\vartheta_1\mathscr D_B\vartheta_1\\
        &-\vartheta_1\mathscr D_A\mathscr D_B\vartheta_1-\frac{1}{2}\gamma_{AB} (\mathscr D_C\vartheta_1\mathscr D^C\vartheta_1\\
        &-\vartheta_1\mathscr D^2\vartheta_1)\bigg],
    \end{split}
\end{equation}
and the evolution equation for $\mathscr U^A$
\begin{widetext}
\begin{equation}
    \label{eq-evo-u4}
    \begin{split}
    \dot{\mathscr U}_A=&\frac{N_A}{3}+\frac{1}{6}(\mathscr D^2N_A-\mathscr D_B\mathscr D_AN^B)+\frac{\pd_u(c_{AB}N^B)}{2}+\frac{\mathscr D_B(mc_A^B)}{2}-\frac{\mathscr D_B\dot{\hat e}_A^B}{4}-\mathscr D_A\mathcal M\\
    &-\frac{c_{BC}\mathscr D_Ac^{BC}}{16}+\frac{c_{AB}\mathscr D_Cc^{BC}}{2}+\frac{1}{4}c^{BC}\mathscr D_C\mathscr D_D\mathscr D_{[A}c_{B]}^D-\frac{1}{4}\mathscr D^Bc_{A}^C\mathscr D_D\mathscr D_{[B}c^D_{C]}\\
    &-\frac{1}{4}\mathscr D_Bc^{BC}\mathscr D_D\mathscr D_Cc_A^D+\frac{1}{64}\pd_u\left[ c_A^B\mathscr D_B\left(c_C^Dc^C_D\right) \right]-\frac{1}{8}\mathscr D_B\left( N_A^B c_C^Dc^C_D\right)\\
    &-\frac{3}{64}\pd_u\left( c_C^Dc^C_D\mathscr D_Bc^B_A \right)+\frac{\mb}{6\kappa}\left[ \frac{\vartheta_1\mathscr D_A\mathscr D^2\vartheta_1}{4}-\frac{\mathscr D_A\vartheta_1}{2}\mathscr D^2\vartheta_1+N\mathscr D_A\vartheta_2\right.\\
    &\left.-\frac{\vartheta_2\mathscr D_AN}{2}+\frac{3\pd_u(\vartheta_1^2\mathscr D_Bc_A^B)}{16} \right]-\frac{\ma}{4\kappa}\left[ \hat\epsilon_A{}^C \left( N\mathscr D_BN_C^B+\frac{N_C^B\mathscr D_BN}{2}-\dot N_C^B\mathscr D_B\vartheta_1\right.\right.\\
    &\left.\left. -\frac{\dot N\mathscr D_Bc_C^B}{2}\right)-\hat\epsilon^{BC}\left( N\mathscr D_BN_{AC}-\frac{1}{2}N_{AC}\mathscr D_BN-\dot N_{AC}\mathscr D_B\vartheta_1+\frac{\dot N\mathscr D_Bc_{AC}}{2} \right) \right],
    \end{split}
\end{equation}
\end{widetext}
where one has not substituted Eqs.~\eqref{eq-eevo} and \eqref{eq-fl2.20} into the above one, otherwise the expression would be much more complicated.

%


\end{document}